# Comparing Homogeneous And Inhomogeneous Time Markov Chains For Modelling Degradation In Sewer Pipe Networks

Lisandro A. Jimenez-Roa[a], Tiedo Tinga[a], Tom Heskes[b], Marielle Stoelinga[a,b]

*[a]University of Twente, Enschede, The Netherlands*
*[b]Radboud University Nijmegen, Nijmegen, The Netherlands*

**Abstract**

Sewer pipe systems are essential for social and economic welfare. Managing these systems requires robust predictive models for degradation behaviour. This study focuses on probability-based approaches, particularly Markov chains, for their ability to associate random variables with degradation. Literature predominantly uses homogeneous and inhomogeneous Markov chains for this purpose. However, their effectiveness in sewer pipe degradation modelling is still debatable. Some studies support homogeneous Markov chains, while others challenge their utility. We examine this issue using a large-scale sewer network in the Netherlands, incorporating historical inspection data. We model degradation with homogeneous discrete and continuous time Markov chains, and inhomogeneous-time Markov chains using Gompertz, Weibull, Log-Logistic and Log-Normal density functions. Our analysis suggests that, despite their higher computational requirements, inhomogeneous-time Markov chains are more appropriate for modelling the nonlinear stochastic characteristics related to sewer pipe degradation, particularly the Gompertz distribution. However, they pose a risk of over-fitting, necessitating significant improvements in parameter inference processes to effectively address this issue.

*Keywords*: Multi-state reliability analysis, degradation modelling, Markov chain, sewer pipe networks.

## 1. Introduction

Sewer networks are essential for societal and economic welfare but face management challenges such as budget constraints, environmental changes, and complex degradation processes. Predictive tools for degradation are becoming crucial as these systems reach the end of their design life, aiding in efficient maintenance and logistics (Ribalta et al. 2023). Robust models for sewer pipe degradation are critical for balancing maintenance costs and system performance, enabling proactive maintenance, informed decision-making, and strategic planning (Caradot et al. 2017).

Comprehensive reviews categorise sewer pipe degradation models into three main types: physics-based, machine learning (ML)-based, and probabilistic models, each with inherent limitations (Ana and Bauwens 2010; Malek Mohammadi et al. 2019; Hawari et al. 2020; Saddiqi et al. 2023; Zeng et al. 2023).

Physics-based models rely on mathematical relations grounded in physical principles, but struggle with complex behaviours in large-scale systems like sewer networks. ML-based models, recognised for identifying patterns in large datasets, are limited by data quality and completeness, affecting their effectiveness (Noshahri et al. 2021). Despite their applications in diagnostics, ML-based models are generally unsuitable for generating reliable and monotonous degradation curves, limiting their utility in long-term maintenance planning (Rokstad and Ugarelli 2015; Caradot et al. 2018; Kantidakis et al. 2023). Comparisons of various ML models for the assessment of sewer pipe condition are discussed in (Nguyen and Seidu 2022; El Morer, Wittek, and Rausch 2023). Probabilistic models consider degradation factors as random variables and share similar drawbacks with ML-based models, including data quality and completeness.

In this work, we focus on *Markov chains*, which are probabilistic models with the ability to predict future distributions associated with degradation processes. Markov chains have several advantages: (i) they convert condition data into ordinal numbers such as severity levels, commonly used in industry to assess the condition of infrastructure assets (Tran, Lokuge, Setunge, et al. 2022); (ii) capture the stochastic nature of degradation processes in sewer pipes; (iii) their outputs can indicate the proportions of pipes in specific conditions, essential for optimising maintenance planning.

Two primary types of Markov chains, homogeneous and inhomogeneous-time, are prevalent in the literature for modelling degradation in sewer pipe networks. However, the optimal Markov chain type remains debated. Proponents of homogeneous-time Markov chains, such as (Micevski, Kuczera, and Coombes 2002), argue for their sufficiency, while proponents of inhomogeneous-time Markov chains, such as (Egger et al. 2013), question homogeneous-time Markov chains efficacy. This gap is what we cover with this work, since no studies have directly compared these models using the same dataset and discussed their suitability.

Understanding this is crucial for sewer asset managers implementing maintenance strategies, as different assumptions about the degradation model can have distinct implications for maintenance policies.

For this, we employ homogeneous-time Markov chains, and for inhomogeneous chains, we use Gompertz, Weibull, Log-Logistic and Log-Normal functions, commonly used in reliability engineering. Our study, using a large-scale sewer network case study in the Netherlands, evaluates calibration complexity and model performance using cross-validation and various goodness-of-fit metrics. We employ the non-parametric Turnbull estimator for handling the interval-censored data in the inspection dataset, serving as a reference.

**Contributions**. Our key contributions are:
- Presenting evidence that inhomogeneous-time Markov chains, despite their complexity, more effectively model non-linear stochastic behaviours in long-lived assets like sewer networks.
- Exploring alternative distributions, such as Log-Logistic and Log-Normal functions, in sewer network degradation modelling.
- Provide comprehensive formal definitions of the degradation models. Additionally, for calibration, we combine the Metropolis-Hastings (M-H) algorithm with the Sequential Least Squares Programming (SLSQP) algorithm for parameter inference in different Markov chains, a novel approach in this field.
- Our implementation is available at https://gitlab.utwente.nl/fmt/degradation-models/ihctmc.

**Paper outline**. Section 2 describes the methods and materials. Section 3 details the experimental setup and results. Section 4 analyzes the findings. 0 concludes the paper and suggests future research directions.

**Related work**. The literature on sewer pipe degradation modelling identifies two primary types of Markov chains: homogeneous and inhomogeneous (Table 1) Homogeneous-time Markov chains (HTMCs) have *constant* transition probabilities, which means that the probabilities of transitioning between states do not change over time. In contrast, inhomogeneous-time Markov chains (IHTMCs) feature *time-variable* transition probabilities, indicating that the likelihood of state transitions can vary.

From the literature, we observe that HTMCs offer simplicity and computational efficiency, making them easier to analyse. However, they often cannot adequately model nonlinear patterns found in stochastic degradation processes, where assuming constant transition probabilities may be overly simplistic. In contrast, IHTMCs can handle these complexities better by accommodating time-varying transition probabilities. Yet, these chains are computationally intensive and sometimes lack feasible closed-form solutions, complicating their application.

Table 1. Research applying different types of Markov chains to model degradation in sewer pipe systems.

| Type | General observations | References |
| --- | --- | --- |
| **Homogeneous-time Markov Chain** | These studies apply *homogeneous-time* Markov chains, assuming time-invariant transition rates and probabilities. Applications extent to modelling specific failure modes, such as corrosion. | Micevski, Kuczera, and Coombes (2002); Baik, Jeong, and Abraham (2006); Dirksen and Clemens (2008); Timashev and Bushinskaya (2015); Lin, Yuan, and Tovilla (2019); Tran, Lokuge, Karunasena, et al. (2022); Jimenez-Roa et al. (2022) |
| **Inhomogeneous-time Markov Chain** | These studies apply *inhomogeneous-time* Markov chains, which model the transition rates based on *survival* functions. Assuming time-variant transition rates and probabilities. | Le Gat (2008); Scheidegger et al. (2011); Egger et al. (2013) |
| **Others** | These studies focus on other forms of Markov chains, such as *semi-Markov chains*, *fuzzy Markov chains*, and *ordered logistic models*. These types of Markov chains are outside the scope of our analysis. | Kleiner (2001); Kleiner, Sadiq, and Rajani (2004); Lubini and Fuamba (2011) |

## 2. Methods and materials

Degradation models for sewer pipes are typically developed using inspection data adhering to standards such as EN13508-1 (2012) and EN13508-2 (2012) }. These standards guide the classification of damages observed through Closed Circuit Television (CCTV) inspections into *severity levels*.

The nature of this data positions these degradation models within the domain of Multi-State Modelling (MSM). MSM captures a system's or component's degradation through *finite states*, associating well-defined degradation indicators for each state, providing a more granular view of the degradation process (Compare et al. 2017).

This is the main reason why the modelling of stochastic degradation of sewer pipes is conducted via Markov chains, as the states in the Markov chain correspond to a *severity level*.

*2.1. Inhomogeneous, homogeneous, continuous, and discrete-time Markov chains*

We start by establishing general definitions and subsequently extracting specific instances. An inhomogeneous continuous-time Markov chain (IHCTMC) is defined by the stochastic process $(X_t)_{t\geq 0}$ (with $t \in [0,\infty)$ representing time) as a tuple $\mathcal{M} = \langle \Omega, S^0, Q(t) \rangle$, where $\Omega$ represents a set of finite states, $S^0$ is an initial state distribution on $\Omega$ where $\sum_{k \in \Omega} S_k = 1$, and $Q(t): \Omega \times \Omega \to \mathbb{R}$ is a time-dependent transition rate matrix. This matrix includes non-diagonal entries $q_{ij}(t)$ for $i, j \in \Omega$ and $i \neq j$, denoting the rate of transitioning from state $i$ to state $j$ at time $t$. The diagonal entries $q_{ii}(t)$ ensure the row sum of $Q(t)$ is zero, reflecting that the rate out of any state equals the sum of the rates into other states. $Q(t)$ can be parameterized by hazard rates $\lambda(t;\theta)$, based on the probability density function $f(t;\theta)$ and the survival function $\mathbf{S}(t;\theta)$, where $\theta$ signifies the function's hyperparameters. The IHCTMC's temporal evolution is described by the *Forward Kolmogorov* equation:

$$\frac{\partial P_{ij}(t,\tau)}{\partial t} = \sum_{k \in \Omega} P_{ik}(t,\tau) Q_{kj}(t) \qquad \text{Eq. (1)}$$

$P_{ij}(t,\tau): \Omega \times \Omega \to [0,1]$ represents the *transition probability matrix*, a continuous and differentiable function detailing the probability of transitioning from state $i$ to state $j$ within the time interval $t$ to $\tau$, with $\tau \geq t$. Using Eq. 1, the *master equation* of the Markov chain is derived, which characterizes the probability flow between states by incorporating *inflow* and *outflow* terms:

$$\frac{\partial S_k(t)}{\partial t} = \sum_{i \in \Omega, i \neq k} S_i(t) Q_{ik}(t) - S_k(t) \left( \sum_{j \in \Omega, j \neq k} Q_{kj}(t) \right) \qquad \text{Eq. (2)}$$

Here $S_k(t)$ is the probability of *being* in state $k \in \Omega$ at time $t$. The term $\sum_{i \in \Omega, j \neq k} Q_{kj}(t)$ captures the rates of transition from state $k$ to all the other states $j$, excluding self-transitions. When the hazard rate $\lambda(t;\theta)$ is assumed to follow an *exponential distribution*, it becomes constant over time, hence time invariant. This introduces the *memoryless* property to the Markov chain formulation, resulting in what is formally known as a homogeneous-continuous-time Markov chain, leading to what is known as *homogeneous continuous-time Markov chain* (HCTMC). Consequently, in Eq. (1) and (2) $Q(t)$ simplifies to $Q$.

If we discretize the time $t$ into discrete intervals of length $\Delta t$, denoted as $n$, we transition from the continuous to the discrete time. It is possible to derive $P$ from $Q$ using the matrix of exponents through the expression $P = e^{-Q\Delta t}$. This approach enables the formulation of the homogeneous discrete-time Markov chain (HDTMC). This form is the simplest among Markov chain models, and state probabilities can be calculated with the Chapman-Kolmogorov equation:

$$S^n = S^0 P^n \qquad \text{Eq. (3)}$$

Here, $S^n$ represents the probability distribution at the $n$th step, and $P^n$ is the $n$-th power of the transition probability matrix. Additional details on Markov chains can be found in Brémaud and Brémaud (2020); Colombo, Abreu, and Martins (2021).

*2.2. Multi-state degradation modelling for sewer networks using parameterized Markov chains*

We first introduce our Markov chain model's structure (Figure 1) and then its parametrisation. Let a pipe element defined with $K$ sequentially arranged states $S = [S_1, S_2, \ldots, S_K]$, where $S_1$ indicates the pristine state and $S_K$, the most deteriorated state, reflects the worst condition for a specific damage type. Considering that inspection data for sewer networks report severities ranging from 1 to 5, and occasionally functional failures, we set $K = 6$.

The transitions in our Markov chain, detailed in Figure 1, allow only progression from better to worse states. All severity levels may progress to functional failure $k = F$. This structure is applicable to IHCTMC, HCTMC, and HDTMC.

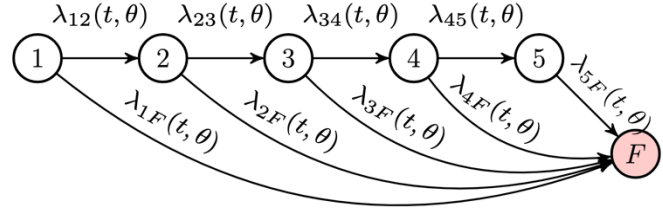

Figure 1 Markov chain structure modelling the degradation of a sewer pipe considering five degradation states and a functional failure state.

The parametrization of our Markov chains involves using probability density functions to model hazard rates. Specifically, we apply Exponential, Gompertz, Weibull, Log-Logistic, and Log-Normal density functions. For the Log-Normal function, which lacks a closed-form hazard rate, we calculate it through the ratio $f(t;\theta)/\mathbf{S}(t;\theta)$. The hazard rates and hyper-parameter ranges for the other functions are detailed in Eq. 4.

| | | | |
|---|---|---|---|
| Exponential | $\lambda(t;\alpha) = \alpha$ | Rate: $\alpha > 0$ | Eq. (4.a) |
| Gompertz | $\lambda(t;\alpha,\beta) = \alpha\beta e^{\beta t}$ | Shape: $\alpha > 0$; Scale: $\beta > 0$ | Eq. (4.b) |
| Weibull | $\lambda(t;\alpha,\beta) = \dfrac{\beta}{\alpha}\left(\dfrac{t}{\alpha}\right)^{\beta-1}$ | Scale: $\alpha > 0$; Shape: $\beta > 0$ | Eq. (4.c) |
| Log-Logistic | $\lambda(t;\alpha,\beta) = \dfrac{(\beta/\alpha)(t/\alpha)^{\beta-1}}{1 + (t/\alpha)^{\beta}}$ | Scale: $\alpha > 0$; Shape: $\beta > 0$ | Eq. (4.d) |

From Eq. 2 we derive the system of differential equations associated with the Markov chain in Figure 1 and present them in Eq. 5.

$$\frac{dS_1(t)}{dt} = \left(-\lambda_{12}(t|\theta) - \lambda_{1F}(t|\theta)\right)S_1(t) \qquad \text{Eq (5.a)}$$

$$\frac{dS_2(t)}{dt} = \lambda_{12}(t|\theta)S_1(t) + \left(-\lambda_{23}(t|\theta) - \lambda_{2F}(t|\theta)\right)S_2(t) \qquad \text{Eq (5.b)}$$

$$\frac{dS_3(t)}{dt} = \lambda_{23}(t|\theta)S_2(t) + \left(-\lambda_{34}(t|\theta) - \lambda_{3F}(t|\theta)\right)S_3(t) \qquad \text{Eq (5.c)}$$

$$\frac{dS_4(t)}{dt} = \lambda_{34}(t|\theta)S_3(t) + \left(-\lambda_{45}(t|\theta) - \lambda_{4F}(t|\theta)\right)S_4(t) \qquad \text{Eq (5.d)}$$

$$\frac{dS_5(t)}{dt} = \lambda_{45}(t|\theta)S_4(t) - \lambda_{5F}(t|\theta)S_5(t) \qquad \text{Eq (5.e)}$$

$$\frac{dS_F(t)}{dt} = \lambda_{1F}(t|\theta)S_1(t) + \lambda_{2F}(t|\theta)S_2(t) + \lambda_{3F}(t|\theta)S_3(t) + \lambda_{4F}(t|\theta)S_4(t) + \lambda_{5F}(t|\theta)S_5(t) \qquad \text{Eq (5.f)}$$

To solve the system of differential equations in Eq. 5, we use the *solve_ivp* function from Python's *scipy.integrate* module. This function, based on `LSODA' from the FORTRAN *odepack* library, solves systems of ordinary differential equations. It utilizes the Adams/BDF method with automatic stiffness detection (SciPy_Community 2023).

*2.3. Model calibration*

To optimise the hyper-parameters of parameterized Markov chains, we employ a novel approach that combines the Metropolis-Hastings (M-H) algorithm (Hastings 1970)—a Markov chain Monte Carlo method—with the Sequential Least Squares Programming (SLSQP) algorithm (Virtanen et al. 2020), specialised in solving constrained nonlinear problems. Although the M-H algorithm alone does not ensure optimal hyper-parameters, it provides a crucial initial guess that aids the SLSQP algorithm in avoiding premature convergence to local optima. To the best of our knowledge, this is the first time these two algorithms have been used together for this application.

Sewer inspections are considered *interval-censored*, where state transitions occur within certain intervals but are not exactly known (Duchesne et al. 2013). This complexity is omitted from our likelihood function, but its further exploration is suggested (Van Den Hout 2016). We analyse the impact of interval-censored data using a non-parametric Turnbull estimator (see Section 2.4).

The initial part of our optimisation problem aligns with Micevski, Kuczera, and Coombes (2002), starting with model calibration in a Bayesian optimization context. We consider $y = [y_1, ..., y_n]$, representing the ages of pipes at inspection. Our likelihood function, $f(y|\gamma, \mathcal{M})$, where $\gamma = \langle \theta, S^0 \rangle$, evaluates the probability of observing the data $y$ given the parameters $\gamma$ and assuming the Markov model $\mathcal{M}$. Incorporating $S^0$ in the optimization introduces the constraint $\sum_{k \in \Omega} S_k^0 = 1$.

Initially, parameters $\gamma$ are sampled from the prior $p(\gamma|\mathcal{M})$. By applying Bayes' theorem, the posterior distribution $p(\gamma|y, \mathcal{M})$ is expressed as:

$$p(\gamma|y, \mathcal{M}) = \frac{f(y|\gamma, \mathcal{M})p(\gamma|\mathcal{M})}{p(y|\mathcal{M})},$$

where the posterior $p(\gamma|y, \mathcal{M})$ updates beliefs about the parameters after observing data. The marginal likelihood $p(y|\mathcal{M})$ is given by:

$$p(\gamma|\mathcal{M}) = \int f(y|\gamma, \mathcal{M})p(\gamma|\mathcal{M})d\gamma,$$

reflecting how well model $\mathcal{M}$, across all parameter values, explains the observed data. Since the computation of $p(y|\mathcal{M})$ is complex, it is assumed that the posterior is *proportional* to the product of the likelihood and the prior.

$$p(\gamma|y, \mathcal{M}) \propto f(y|\gamma, \mathcal{M})p(\gamma|\mathcal{M}) \quad \text{Eq. (6)}$$

For our optimization problem, we first derive the following relations:

$$\mathbf{S}_k(t; \gamma, \mathcal{M}) = \sum_{m=1}^{k} S_m(t; \gamma, \mathcal{M}),$$

$$f(y|\gamma, \mathcal{M}) = -\frac{d\mathbf{S}_k(t; \gamma, \mathcal{M})}{dt},$$

where $\mathbf{S}_k(t; \gamma, \mathcal{M})$ is the survival functions, notice that $\mathbf{S}_{k=F}(t; \gamma, \mathcal{M}) = 1$. Then the log-likelihood function ($\ell$) is defined by:

$$\ell = \sum_{t \in y} \sum_{k \in \Omega} \mathbf{n}_{\underline{k},t} \log\left(-\frac{d\mathbf{S}_k(t; \gamma, \mathcal{M})}{dt}\right) \quad \text{Eq. (7)}$$

Here, $\ell \in (-\infty, 0]$, $\mathbf{n}_{\underline{k},t}$ denotes the number of pipes of age $t$ found in states that transitioned from $k$, denoted as $\underline{k}$. E.g., if $k = 1$, $\underline{k} = \langle 2, F \rangle$ (see Figure 1).

The acceptance distribution $\mathcal{A}$ of the M-H algorithm is given by:

$$\mathcal{A}(x_t, x_{t+1}) = \min\left(1, \frac{f(y|\gamma_{t+1}, \mathcal{M})p(\gamma_{t+1}|\mathcal{M})}{f(y|\gamma_t, \mathcal{M})p(\gamma_t|\mathcal{M})}\right) > U(0,1) \quad \text{Eq. (8)}$$

Here, $x_t$ and $x_{t+1}$ are the current and proposed points in the parameter space, and $\gamma_t$ and $\gamma_{t+1}$ the corresponding sets of hyper-parameters. The prior $p(\gamma_{t+1}|\mathcal{M})$ is a uniform distribution $U(\underline{\epsilon}, \overline{\epsilon})$, where $\underline{\epsilon}$ and $\overline{\epsilon}$ define the range for each hyper-parameter in $\gamma$.

The M-H algorithm executes 50,000 iterations, with the first 49,000 as the burn-in period, and the subsequent 1,000 samples used to compute mean values and the output $\gamma_{M-H}$.

Post convergence, $\gamma_{M-H}$ serves as the initial guess for SLSQP, with $\gamma$ parameters constrained between $\underline{\epsilon}$ and $\overline{\epsilon}$. SLSQP employs convergence tolerances of *eps = 1E-5* and *ftol = 1E-50*, and runs for up to 300 iterations. Upon SLSQP convergence, $\gamma_{SLSQP}$ is derived and selected as the optimal set of hyper-parameters for further analysis.

*2.4. Non-parametric*

Non-parametric survival curve estimators compute survival probabilities without assuming a specific distribution for survival times. This approach provides a reliable baseline crucial in our analysis to understand the effects of interval-censored data. Given the interval-censored nature of our data, we employ the *Turnbull* estimator (Turnbull 1976), a non-parametric technique suitable for such data.

For each severity level $k$ in our Markov chain, we calibrate a Turnbull estimator. Data binarization is achieved using $k_{bin} \in \Omega$ as a threshold. Observations with $k < k_{bin}$ are considered *non-events*, associated with the interval $[y_i, +\infty)$. Conversely, $k \geq k_{bin}$ are treated as *events*, defined by the interval $[0, y_i)$. These non-parametric Turnbull estimators are computed using the *lifelines* toolbox (Davidson-Pilon 2019) in Python.

*2.5. Goodness-of-fit metrics*

Markov chains performance is evaluated via likelihood-based metrics: Akaike Information Criterion (AIC) (Akaike 1998) (Eq. (9.a)) and Bayesian Information Criterion (BIC) (Schwarz 1978) (Eq. (9.b)), which aid in model selection. Moreover, the Root Mean Squared Error (RMSE) (Eq. (9.c)) quantifies the Euclidean distance between the predictions of the Markov chains and the Turnbull estimator.

$$\text{AIC} = 2|\gamma| - 2\ell \qquad \text{Eq. (9.a)}$$

$$\text{BIC} = \ln(|y|)|\gamma| - 2\ell \qquad \text{Eq. (9.b)}$$

$$\text{RMSE} = \sqrt{\frac{1}{|y| \times K} \sum_{t \in y} \sum_{k \in \Omega} \left(S_k(t) - \hat{S}_k(t)\right)^2} \qquad \text{Eq. (9.c)}$$

Both AIC and BIC include $\gamma$, the number of parameters in the model, with BIC additionally considering $\ln(|y|)$, the natural logarithm of the sample size. RMSE involves $S_k(t)$ and $\hat{S}_k(t)$, which denote the probabilities of being in state $k$ at pipe age $t$, obtained from the Markov chains and the Turnbull estimator, respectively.

## 3. Experimental setup and evaluation

*3.1. Case study*

Our case study focuses on a sewer pipe network in Breda, Netherlands, using inspection data from 1940 to 2020. During inspections, damage type and severity are recorded according to European norms (EN13508-1 2012; EN13508-2 2012). Malek Mohammadi et al. (2020); Salihu et al. (2022) identify the primary factors affecting sewer pipe condition as age, material, and content. Based on these, we categorize pipes into three cohorts and examine the damage code *BAF*, indicative of infiltration:
-   Cohort **CMW**: Concrete pipes carrying mixed and waste content, Length: 469 km, Pipes: 11,942.
-   Cohort **CS**: Concrete pipes carrying storm water, Length: 172 km, Pipes: 4,701.
-   Cohort **PMW**: PVC pipes carrying mixed and waste content, Length: 294 km, Pipes: 10,777.

*3.2. Experimental setup*

Our experiment aims to assess the efficacy of homogeneous and inhomogeneous Markov chains in predicting stochastic degradation of sewer pipes using the same dataset. Employing cross-validation, 70% of the sewer pipes from the case study are randomly selected for model calibration, while the remaining 30% is used to compute goodness-of-fit metrics described in Section 2.5.

*3.3. Results*

The different types of Markov chains are calibrated using data from cohorts **CMW**, **CS**, and **PMW** on infiltration, following the procedure described in Section 2.3 using the training set.
Table 2 presents the goodness-of-fit metrics for both the training and test sets, while Figure 2 illustrates the state probabilities. The results from the Turnbull estimator for both sets are also displayed. The vertical grey dashed lines in the figures denote the last inspection used for model training.
By solving Eq. (1), we obtain the transition probability matrix over time $P_{i,j}(t, \tau)$. Figure 3 displays these probabilities for cohort **CS** and infiltration.

## 4. Findings

*4.1. Comparison between cohorts*

For all cohorts, the inhomogeneous time Markov chains (modeled with Gompertz, Weibull, Log-logistic, and Log-normal functions) outperform the homogeneous time Markov chains (modeled with the Exponential function and HDTMC) by achieving the lowest values in all goodness-of-fit metrics in Table 2.

Table 2 Goodness-of-fit metrics computed on the training and testing sets for different Markov chains. Blue and red colors indicate the best and worst scores, respectively.

| Cohort | Type | Func. | $|\gamma|$ | Training set | | | Test set | | |
|---|---|---|---|---|---|---|---|---|---|
| | | | | RMSE | AIC | BIC | RMSE | AIC | BIC |
| **CMW** | IHCTMC | Gompertz | 24 | 0.025 | 57431 | 57601.8 | **0.0337** | **20436** | **20583.2** |
| | IHCTMC | Weibull | 24 | 0.0233 | **57414.8** | **57585.5** | 0.0361 | 20478 | 20625.2 |
| | IHCTMC | Log-Logistic | 24 | **0.0219** | 58544.4 | 58715.1 | **0.0391** | 20861.2 | 21008.4 |
| | IHCTMC | Log-Normal | 24 | 0.0221 | 58553.6 | 58724.3 | 0.0385 | 20823.2 | 20970.3 |
| | HCTMC | Exponential | 15 | **0.0312** | **59574.6** | **59681.3** | 0.0359 | **21142.1** | **21234** |
| | HDTMC | - | 15 | **0.0312** | **59574.6** | **59681.3** | 0.0359 | **21142.1** | **21234** |
| **CS** | IHCTMC | Gompertz | 24 | 0.0358 | **3532.8** | **3665.8** | 0.0468 | **1179.4** | **1290.9** |
| | IHCTMC | Weibull | 24 | 0.0326 | 3850 | 3983 | **0.0423** | 1269 | 1380.5 |
| | IHCTMC | Log-Logistic | 24 | **0.0313** | **4111.3** | **4244.3** | 0.0427 | 1345.9 | **1457.4** |
| | IHCTMC | Log-Normal | 24 | 0.033 | 4035.1 | 4168.1 | 0.0446 | 1324.4 | 1436 |
| | HCTMC | Exponential | 15 | **0.0593** | 4006.7 | 4089.8 | **0.0583** | 1359.1 | 1428.8 |
| | HDTMC | - | 15 | **0.0593** | 4006.7 | 4089.8 | **0.0583** | 1359.1 | 1428.8 |
| **PMW** | IHCTMC | Gompertz | 24 | 0.0199 | **2349.3** | **2495.3** | 0.0172 | **989.7** | 1114.7 |
| | IHCTMC | Weibull | 24 | **0.0153** | **5522** | **5668** | 0.0403 | **1822.1** | **1947** |
| | IHCTMC | Log-Logistic | 24 | 0.0217 | 4699.7 | 4845.7 | 0.0178 | 1523.1 | 1648 |
| | IHCTMC | Log-Normal | 24 | 0.0211 | 3588.2 | 3734.2 | 0.0179 | 1493.6 | 1618.6 |
| | HCTMC | Exponential | 15 | **0.0297** | 2438.6 | 2529.9 | 0.0256 | 1002.8 | **1080.9** |
| | HDTMC | - | 15 | **0.0297** | 2438.6 | 2529.9 | 0.0256 | 1002.8 | **1080.9** |

Notice that a smaller RMSE in Table 2 suggests a closer alignment of the Markov chains with the Turnbull estimator. Also, the goodness-of-fit metrics for both homogeneous time Markov chains are identical, which is consistent with the theoretical mapping of one into the other. This is visually corroborated in Figure 2.

In Table 3, we compute the error for each goodness-of-fit metric in Table 2 between the best performing inhomogeneous and homogeneous Markov chain for each cohort, relative to the largest value.

From Table 3, it is evident that inhomogeneous Markov chains generally improve over homogeneous Markov chains, except for the BIC results for cohort **PMW** (marked with †), where the homogeneous Markov chains showed more favorable outcomes.

Table 3. Error between the best inhomogeneous and homogeneous Markov chains.

| Cohort | Training set | | | Test set | | |
|---|---|---|---|---|---|---|
| | RMSE | AIC | BIC | RMSE | AIC | BIC |
| CMW | 29.8% | 3.6% | 3.5% | 6.1% | 3.3% | 3.1% |
| CS | 47.2% | 11.8% | 10.4% | 27.4% | 13.2% | 9.7% |
| PMW | 48.5% | 3.7% | 1.4% | 32.8% | 1.3% | 3.0%† |

*4.2. Transition probabilities over time*

For further clarification and illustrative purposes in understanding the behaviour within different types of Markov chains, Figure 3 displays the transition probability variations among Markov chains in the **CS** cohort. The homogeneous time Markov chain, employing the Exponential distribution, maintains constant transition probabilities over time, reflecting its homogeneous and memoryless properties. Conversely, the inhomogeneous time Markov chains reveal diverse behaviours in their transition probabilities, depicting distinct temporal variations. Notice that there are also differences in the transition probabilities between inhomogeneous Markov chains, due to the different assumptions on the density functions.

*4.3. Overfitting*

All inhomogeneous Markov chains map well where data is available (up to around 70-year-old pipes, see gray dashed vertical lines in Figure 2), however, beyond this point, these models tend to move faster to worse conditions. This is likely related to the additional degrees of freedom that inhomogeneous Markov provides.

This effect is less in homogeneous Markov chains because they have fewer degrees of freedom. Thus, future research should consider this aspect in the model calibration, to improve the predictive capabilities of inhomogeneous time Markov chains.

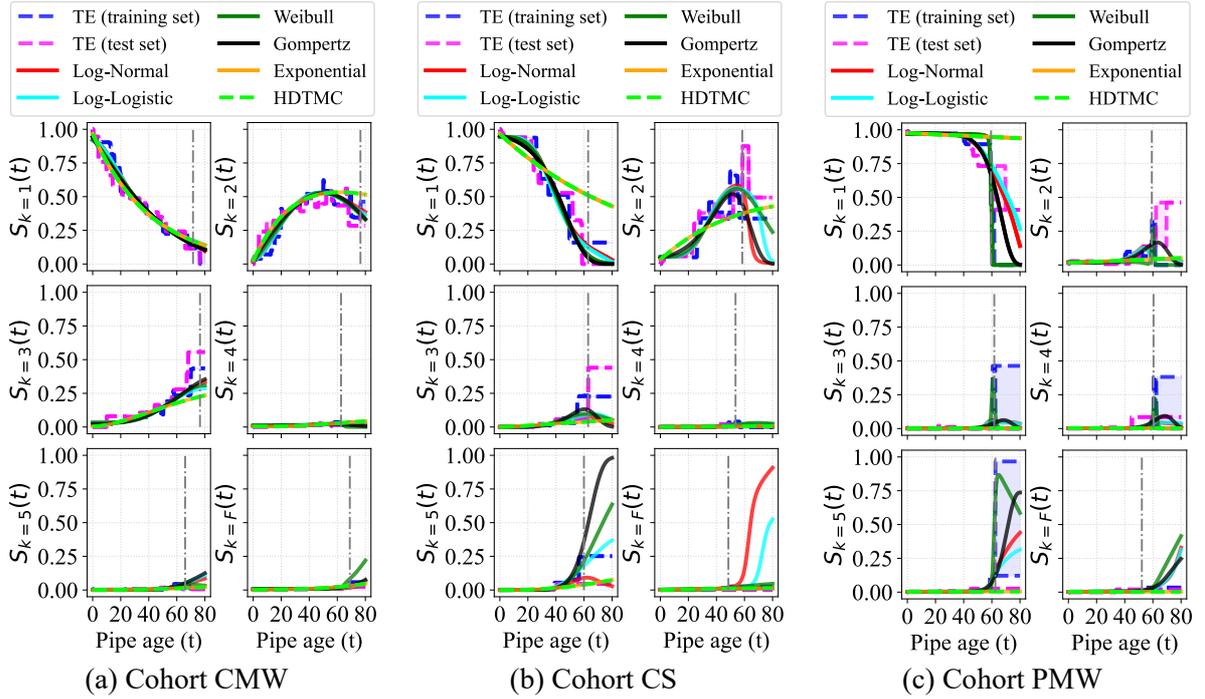

Figure 2 State probability $S_k(t)$ for different Markov chains. Dashed lines are the Turnbull estimators. For Cohort (a) **CMW**, (b) **CS**, (c) **PMW**.

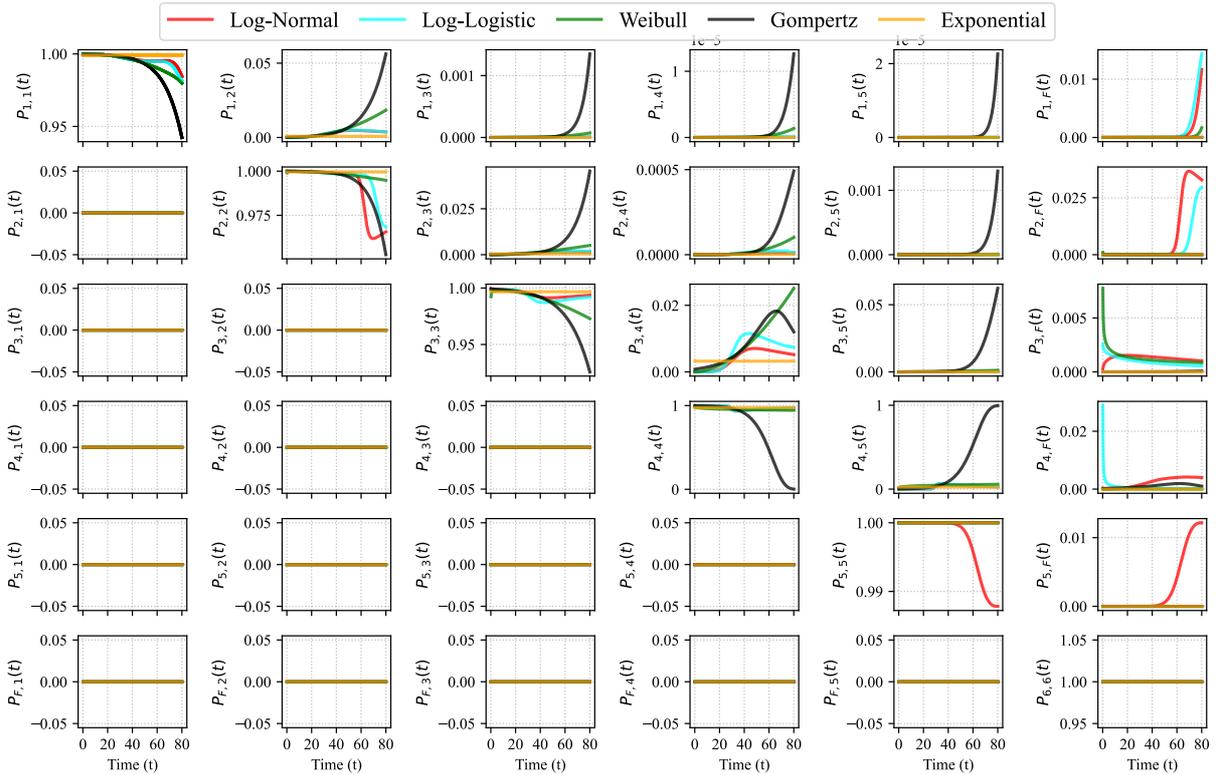

Figure 3 Transition probabilities $P_{i,j}(t, \tau)$ for cohort **CS** on infiltration.

### 4.4. Comparing inhomogeneous Markov chains

Upon closer examination of inhomogeneous Markov chains modeled with Log-Normal, Log-Logistic, Weibull, and Gompertz density functions, Table 2 reveals that the Gompertz distribution consistently demonstrates good performance across all cohorts and goodness-of-fit metrics, followed by Weibull and Log-Logistic density functions. Notably, the Weibull distribution shows poor performance for cohort PMW, likely due to sub-optimal parameters resulting from convergence in local optima.

## 5. Conclusions and future research

We examine the effectiveness of homogeneous and inhomogeneous Markov chains in modelling stochastic degradation in sewer pipes. We introduce four inhomogeneous Markov chain models parameterized with Log-Normal, Log-Logistic, Weibull, and Gompertz density functions, and compare them against a homogeneous Markov chain with an Exponential distribution and discrete-time Markov chains using the same dataset.

These models are calibrated using Metropolis-Hastings and Sequential Least Squares Programming algorithms, utilizing historical inspection data from a Dutch sewer network. Additionally, we employ the Turnbull estimator as a reference to account for the interval-censoring in the dataset.

From the dataset, we establish three cohorts and assess the fit of the Markov chains using various goodness-of-fit metrics. Our findings suggest that, despite their complexity, inhomogeneous time Markov chains more effectively model the nonlinear stochastic behaviours observed in sewer network inspection data. In particular, the inhomogeneous time Markov chain modeled with the Gompertz distribution consistently showed good performance.

This observation aligns with Mizutani and Yuan (2023), which recommends inhomogeneous Markov chains to model time-varying transition probabilities in bridge structures. This result is crucial for sewer asset managers, as deriving maintenance policies for sewer pipes requires accounting for these nonlinearities in degradation models, since different assumptions may yield distinct maintenance policy implications.

To maintain the severity levels within the model and to address the nonlinearities in the degradation process of sewer pipes, it is essential to adequately evaluate the inhomogeneous behaviours. The use of homogeneous time Markov chains is advised only if the modeler can substantiate this assumption beforehand.

**Future research**. Future research directions include:
- Addressing the omission of interval censoring during the calibration of our inhomogeneous time Markov chains, which approximate the Turnbull estimator, requires further investigation to assess the validity of neglecting interval censoring.
- Expanding our models to consider pipe length and the distribution of degradation along the sewer pipe, beyond focusing solely on the most severe pipe condition during inspections.
- Developing models that incorporate covariates without forming cohorts to minimize cohort selection biases.
- Examining model performance in various cities, delving into domain adaptation using tools like Reinforcement and Transfer Learning.
- Integrating uncertainty quantification, vital for decision making, requires studies on accurate uncertainty bound estimation.
- Despite our calibration process's efficacy, further exploration of alternative optimisation techniques for nonlinear constrained problems is needed to improve parameter inference speed, aiming to reduce over-fitting.
- Future studies should also investigate the application of these models to optimize maintenance and inspection policies in sewer networks.

## Acknowledgements

This research has been partially funded by NWO under the grant PrimaVera (https://primavera-project.com) number NWA.1160.18.238, by the ERC Consolidator grant CAESAR number 864075, and by the European Union's Horizon 2020 research and innovation programme under the Marie Sklodowska-Curie grant agreement No 101008233.